\documentclass[aps,reprint,twocolumn,superscriptaddress,citeautoscript,amsmath,amssymb,floatfix]{revtex4-2}
\usepackage{amsmath}
\usepackage{amssymb}
\usepackage{mathtools}
\usepackage{textcomp,gensymb} %For degree symbol
\usepackage{bm} %For bold math (Greek) symbols or bold italic in math mode
\usepackage[usenames,dvipsnames]{xcolor}
\usepackage{float}
\usepackage{setspace}
\usepackage{dcolumn}
\usepackage{array}

\usepackage{changepage}
\usepackage{tikz}
\usepackage{contour}
\usepackage[left]{lineno}
\usepackage{blindtext}
\usepackage{setspace}
\usepackage[colorlinks=true,citecolor=blue,linkcolor=blue,pdftex,hypertexnames=false,urlcolor=blue,linktocpage=true]{hyperref}
\usepackage[none]{hyphenat}%to remove hyphen 
\usepackage{adjustbox}
\newcolumntype{P}[1]{>{\centering\arraybackslash}p{#1}}
\usepackage{lipsum}

\makeatletter
\newcommand*{\rom}[1]{\expandafter\@slowromancap\romannumeral #1@}
\makeatother

\newcommand{\tildeapprox}{\raisebox{0.5ex}{\texttildelow}}
\newcommand{\ckfa}[0]{$\text{Ca}\text{K}\text{Fe}_4\text{As}_4$}
\newcommand{\ckfna}{CaK(Fe$_{\mathrm{1-}x}$Ni$_x$)$_4$As$_4$}
\newcommand{\ckfmna}{CaK(Fe$_{\mathrm{1-}x}$Mn$_x$)$_4$As$_4$}

\newcommand{\akfa}[0]{$AeA\text{Fe}_4\text{As}_4$}

\newcommand{\bkfmna}{Ba$_{0.5}$K$_{0.5}$Fe$_{\mathrm{1-}x}$Mn$_x$As$_2$}
\newcommand{\lfmaf}{LaFe$_{\mathrm{1-}x}$Mn$_x$As$_{0.89}$F$_{0.11}$}

\newcommand{\bfa}[0]{$\text{Ba}\text{Fe}_2\text{As}_2$}

\newcommand{\bfnca}{Ba(Fe$_{\mathrm{0.9-}x}$Ni$_{0.1}$Cr$_x$)$_2$As$_{2}$}

\newcolumntype{d}[1]{D{.}{.}{#1}}

\begin{document}
\title{Antiferromagnetic order and its interplay with superconductivity in CaK(Fe$_{1-x}$Mn$_x$)$_4$As$_4$}

\author{J.~M.~Wilde}
\affiliation{Ames National Laboratory, U.S. DOE, Iowa State University, Ames, Iowa 50011, USA}
\affiliation{Department of Physics and Astronomy, Iowa State University, Ames, Iowa 50011, USA}

\author{A. Sapkota}
\affiliation{Ames National Laboratory, U.S. DOE, Iowa State University, Ames, Iowa 50011, USA}

\author{Q.-P. Ding}
\affiliation{Ames National Laboratory, U.S. DOE, Iowa State University, Ames, Iowa 50011, USA}
\affiliation{Department of Physics and Astronomy, Iowa State University, Ames, Iowa 50011, USA}

\author{M. Xu}
\affiliation{Ames National Laboratory, U.S. DOE, Iowa State University, Ames, Iowa 50011, USA}
\affiliation{Department of Physics and Astronomy, Iowa State University, Ames, Iowa 50011, USA}

\author{W. Tian}
\affiliation{Neutron Scattering Division, Oak Ridge National Laboratory, Oak Ridge, Tennessee 37831, USA}

\author{S.~L.~Bud'ko}
\affiliation{Ames National Laboratory, U.S. DOE, Iowa State University, Ames, Iowa 50011, USA}
\affiliation{Department of Physics and Astronomy, Iowa State University, Ames, Iowa 50011, USA}

\author{Y. Furukawa}
\affiliation{Ames National Laboratory, U.S. DOE, Iowa State University, Ames, Iowa 50011, USA}
\affiliation{Department of Physics and Astronomy, Iowa State University, Ames, Iowa 50011, USA}

\author{A. Kreyssig}
\affiliation{Ames National Laboratory, U.S. DOE, Iowa State University, Ames, Iowa 50011, USA}
\affiliation{Department of Physics and Astronomy, Iowa State University, Ames, Iowa 50011, USA}
\affiliation{Institute for Experimental Physics \rom{4}, Ruhr-Universit\"{a}t Bochum, 44801 Bochum, Germany}

\author{P.~C.~Canfield}
\affiliation{Ames National Laboratory, U.S. DOE, Iowa State University, Ames, Iowa 50011, USA}
\affiliation{Department of Physics and Astronomy, Iowa State University, Ames, Iowa 50011, USA}

\date{\today}

\begin{abstract}
The magnetic order for several compositions of CaK(Fe$_{1-x}$Mn$_x$)$_4$As$_4$ has been studied by nuclear magnetic resonance (NMR), M\"ossbauer spectroscopy, and neutron diffraction. Our observations for the Mn-doped 1144 compound are consistent with the hedgehog spin vortex crystal (hSVC) order which has previously been found for Ni-doped $\text{Ca}\text{K}\text{Fe}_4\text{As}_4$. The hSVC state is characterized by the stripe-type propagation vectors $(\pi\,0)$ and $(0\,\pi)$ just as in the doped 122 compounds. The hSVC state preserves tetragonal symmetry at the Fe site, and only this SVC motif with simple AFM stacking along $\textbf{c}$ is consistent with all our observations using NMR, M\"ossbauer spectroscopy, and neutron diffraction. We find that the hSVC state in the Mn-doped 1144 compound coexists with superconductivity (SC), and by combining the neutron scattering and M\"ossbauer spectroscopy data we can infer a quantum phase transition, hidden under the superconducting dome, associated with the suppression of the AFM transition temperature ($T_\text{N}$) to zero for $x\approx0.01$. In addition, unlike several 122 compounds and Ni-doped 1144, the ordered magnetic moment is not observed to decrease at temperatures below the superconducting transition temperature ($T_\text{c}$). 

\end{abstract}

\maketitle

\section{Introduction}

An intriguing aspect of high-$T_\mathrm{c}$ iron--based superconductors (IBSC) are their phase diagrams showing the intricate interrelationships of easily tunable structural, magnetic and superconducting ground states \cite{Canfield2010feas,paglione2010high}. Systematic studies of these phase diagrams have provided insights into these different orders, and the interrelationships between them \cite{Canfield2010feas,Kreisel_2020}. Discovery of 1144 \akfa{}  ($Ae$~=~Ca, Sr; $A$~=~K, Rb, Cs) \cite{Iyo2016new} opened a new avenue of research to revisit these different orders, and the fundamentals of high-$T_\mathrm{c}$ superconductivity (SC) in IBSC. A recent addendum to interesting aspects of this family is the prediction of topological behavior in the form of magnetic-field-induced Weyl points in \ckfa{} \cite{Heinsdorf_2021_weyl}.

\begin{figure}[!htb]
    \centering
    \includegraphics[width=\columnwidth]{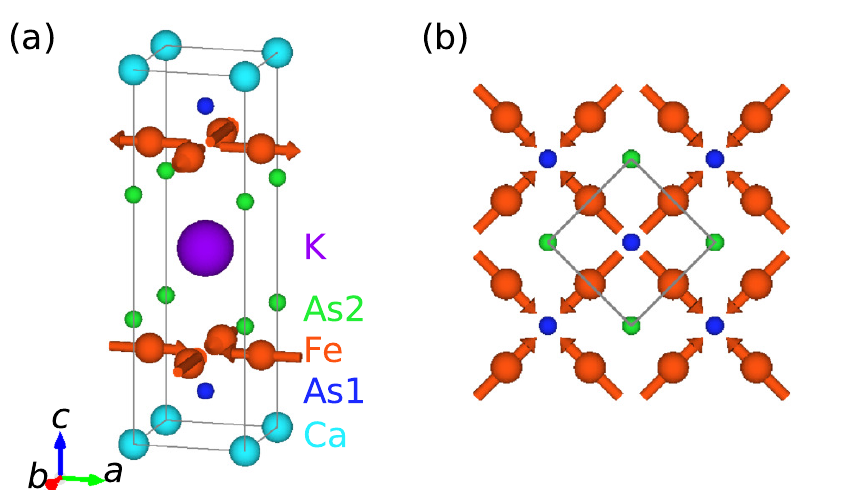}
    \caption{Chemical and magnetic structure of $\text{CaKFe}_4\text{As}_4$. (a) Magnetic structure of $\text{CaKFe}_4\text{As}_4$ with red arrows indicating magnetic moments consistent with hSVC AFM order for one chemical unit cell with space group $P4/mmm$ (b) hSVC AFM order in one As-Fe-As layer. Solid grey lines show the chemical unit cell.}
    \label{CSMO}
\end{figure}

1144 compounds are closely related to the 122 iron pnictides, $Ae\text{Fe}_2\text{As}_2$, but are different in several ways. First, unlike $Ae\text{Fe}_2\text{As}_2$, stoichiometric \akfa{} is a superconductor with transition temperature $T_{\text{c}}=$~35~K, and manifests no other magnetic or structural phase transitions below 300~K \cite{Iyo2016new,Meier2016anisotropic}. This provides a unique opportunity to study high-$T_\mathrm{c}$ SC without impurity/disorder effects from substitution. Another prominent difference from 122 compounds is the overall reduction in crystallographic symmetry. Alternating Ca and K layers, as shown in Fig.~\ref{CSMO}(a), break the body centering and results in inequivalent As1 and As2 sites. Also, the point symmetry of the Fe site is now orthorhombic instead of tetragonal, which has enormous effects on its magnetism \cite{Meier2018,Kreyssig1144_2018}. \ckfa{} can be considered to be electronically analogous to near optimally hole-doped BaFe$_2$As$_2$, Ba$_{0.5}$K$_{0.5}$Fe$_2$As$_2$, and shows similar features in thermodynamic and transport measurements \cite{Meier2016anisotropic,Meier2018}. The similiarity to  Ba$_{0.5}$K$_{0.5}$Fe$_2$As$_2$ motivated the tuning of the physical properties of \ckfa{} using chemical substitution to potentially induce magnetic order.
%; the lower symmetry favors experimentally unrealized spin-vortex crystal (SVC) order \cite{Meier2018}. 

Indeed, electron doping by using Co or Ni substitution at the Fe site successfully suppresses SC and induces magnetic order which coexists microscopically with SC \cite{Meier2018}. However, the magnetic order is different from the often observed stripe-type spin-density wave (SSDW) order with orthorhombic symmetry observed in the 122 compounds. Results from NMR and M\"ossbauer spectroscopy combined with a careful symmetry analysis proposed a new form of non-colinear hedgehog spin-vortex crystal (hSVC) magnetic order with tetragonal symmetry in the Fe planes, as illustrated in Fig.~\ref{CSMO}(b) \cite{Meier2018}. The in-plane arrangement of Fe moments are 45\degree\ to those of the colinear SSDW order in the 122 compounds, but are described by same symmetry equivalent propagation vectors $(\pi\,0)$ and $(0\,\pi)$. Neutron diffraction measurements on Ni-doped 1144, discussed in Ref.~\citenum{Kreyssig1144_2018}, later confirmed and provided its detailed nature. hSVC order was found to be long-range, commensurate to the lattice, and with a simple antiferromagnetic (AFM) arrangement of spins along the \textbf{c} direction. Furthermore, the suppression of the magnetic moment below $T_c$, similar to electron-doped 122, suggests the presence of competing and coexisting magnetism and SC \cite{Kreyssig1144_2018,bud2018coexistence}.

The effects of hole doping on the Fe site in the 122 IBSC are demonstrably different from electron doping \cite{Sefat_2009_Cr,bud2009structural,kim2010antiferromagnetic,thaler2011physical,Marty_2011,Tucker_2012,Hammerath_2014}. In particular, Mn/Cr substitution on stoichiometric 122 compounds affects the SSDW and orthorhombic order, but instead of SC, two-$\tau$ AFM order with propagation vectors $(\pi, 0)$ and $(0, \pi)$, competing checkerboard AFM order, and spin fluctuations appear. In addition, Mn/Cr substitution on already superconducting samples, such as \bkfmna{}\cite{Cheng_2010} and \bfnca{}\cite{Gong_2018}, rapidly suppresses the SC and stripe-type AFM order of the parent \bfa{}. To study the effects of Mn on SC and possibl2e magnetic order in 1144 compounds, systematic studies of thermodynamic and transport properties were done on \ckfmna{} as reported in Ref.~{\citenum{Mingyu_2021}}. A detailed temperature-composition ($T-x$) phase diagram which shows the appearance of an AFM phase after some degree of suppression of SC was obtained. Qualitatively, the phase diagram seems to be very similar to the electron-doped 1144. Quantitatively, Mn substitution appears to suppress $T_{\text{c}}$ much more rapidly. This said, several questions regarding the detailed nature of magnetism remain. Is the magnetic order hSVC or checkerboard, long-range or short-range, commensurate or incommensurate? Are there any signs of the interplay between magnetism and SC? 

\begin{figure}[]
	\centering
	\includegraphics[width=\columnwidth]{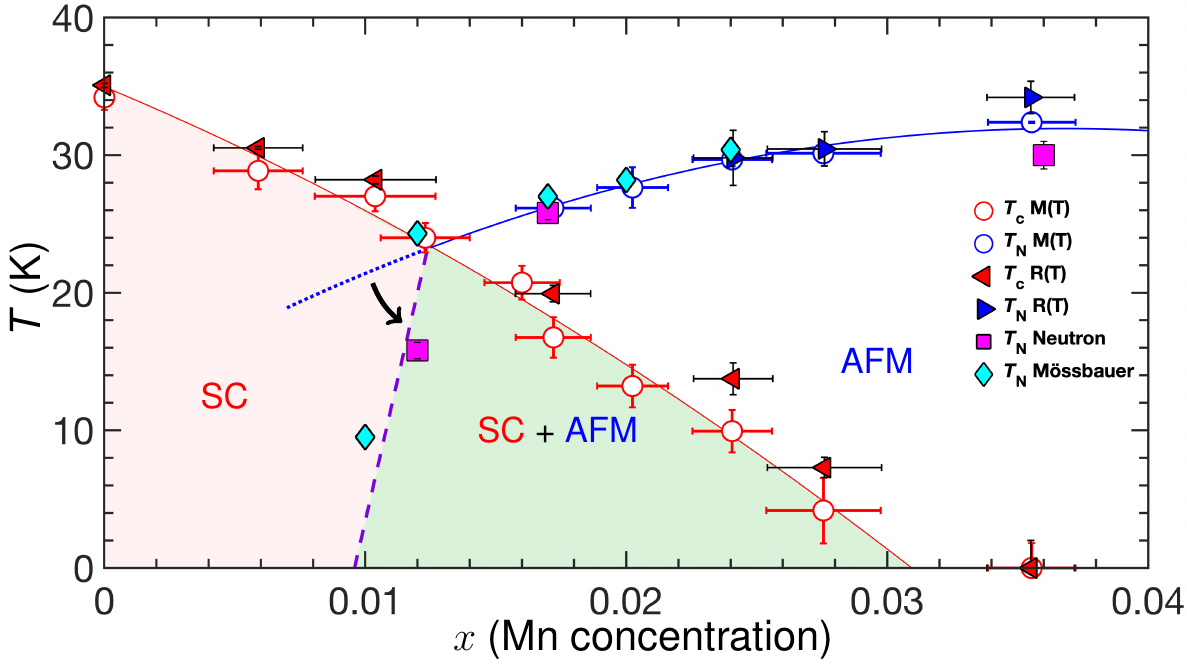}
	\caption{Temperature-composition ($T-x$) phase diagram of \ckfmna\ single crystals. The $T_\mathrm{N}$ from neutron diffraction and M\"ossbauer spectroscopy measurements are added to the resistance [$R(T)$] and magnetization [$M(T)$] results contained in the phase diagram in Ref.~\cite{Mingyu_2021}. The solid red and blue lines corresponding to $T_\mathrm{c}$ and $T_\mathrm{N}$, respectively, are parabolic fits to the respective data obtained from magnetization measurements. Magenta filled squares and cyan filled diamonds are $T_\mathrm{N}$ corresponding to our current neutron diffraction and M\"ossbauer spectroscopy measurements, respectively. The dashed line shows the behavior of $T_\mathrm{N}$ when it is below $T_\mathrm{c}$. These data are uniquely from the M\"ossbauer and neutron diffraction data sets and allow for an estimate of the critical doping value, $x_\mathrm{c}~\tildeapprox{}~0.01$, for the zero temperature onset of AFM ordering under the SC dome.}
	\label{PD}
\end{figure}

We present NMR, M\"ossbauer spectroscopy, and neutron diffraction measurements performed on \ckfmna{} single crystals to address the above questions and understand the magnetic order in detail. We have refined the $T-x$ phase diagram from Ref.~\cite{Mingyu_2021} with additional measurements of $T_{\text{N}}$ including several below $T_{\text{c}}$ as shown in Fig.~\ref{PD}, which shows three low-temperature regions, SC, SC + AFM (coexistence region) and AFM. We found that the AFM phase for the Mn-doped 1144 compound is qualitatively and quantitatively the same hSVC order as the Ni-doped case. As expected the hSVC order is found to be long range, commensurate and non-colinear with AFM arrangement of spins along the \textbf{c}-direction. Whereas, like Ni-doping we infer a dramatic suppression of $T_\mathrm{N}$ for $T_\mathrm{N}<T_\mathrm{c}$, unlike Ni-doping, there is no indication of any suppression of the ordered magnetic moment below $T_\text{c}$. We discuss the possible scenario that similar to \bkfmna{}, Mn dopants in 1144 can act as local magnetic impurities rather than hole dopants. In this case, the suppression of SC is due to local magnetic impurities, which also help stabilize the competing AFM order.

The rest of this paper is organized as follows. Experimental details are given in Sec. II. The results and analysis are presented in Sec. III, where we begin with the NMR measurements followed by neutron diffraction, and M\"ossbauer spectroscopy measurements. A brief discussion and summary is presented in Sec. IV.

\section{Experimental Details}

Single crystals of \ckfmna{} were grown from high-temperature solution rich in transition-metals and arsenic as discussed in detail in Ref.~\cite{Mingyu_2021}. Wavelength-dispersive x-ray spectroscopy employing a JEOL JXA-8200 microprobe system on cleaved surfaces of crystals from the same batch was done for the determination of the compositions. Detail characterization of the physical properties of \ckfmna{} series are presented in Ref.~\cite{Mingyu_2021}. 

NMR spectrum measurements of $^{75}$As ($I$ = $\frac{3}{2}$, $\frac{\gamma_{\rm N}}{2\pi}$ = 7.2919 MHz/T, $Q=$ 0.29 barns) nuclei were conducted using a lab-built phase-coherent spin-echo pulse spectrometer on single crystals of CaK(Fe$_{0.964}$Mn$_{0.036}$)$_4$As$_4$ with $T_{\rm N}$ = 32 K. Five  plate-shaped  single crystals (a typical size of $\sim$2 $\times$ $\sim$2 $\times$ 0.1 mm$^3$ for each) with a total mass of $\sim$ 10 mg were placed on a glass plate  to align their $c$-axis directions. The $^{75}$As-NMR spectra were obtained at a resonance frequency of $f$ = 43.2 MHz by sweeping the magnetic field $H$  in two different states of the paramagnetic state ($T$ = 35 K) and the antiferromagnetic ordered state ($T$ = 1.6 K). 

M\"ossbauer spectroscopy measurements were performed on five Mn concentrations, $x=$~0.010, 0.012, 0.017, 0.020, and 0.024 using a SEE Co. conventional, constant acceleration type spectrometer in transmission geometry with a $^{57}$Co(Rh) source kept at room temperature. The absorbers were prepared as a mosaic of single crystals held on a VWR Weighting Paper disk by a small amount of Apiezon N grease. An effort was made to keep gaps between crystals to a minimum. The $c$-axis of the crystals in the mosaic was parallel to the $\gamma$ beam. The absorber was cooled to a desired temperature using a Janis model SHI-850-5 closed cycle refrigerator (with vibration damping). The driver velocity was calibrated using an $\alpha$-Fe foil, and all isomer shifts (IS) are quoted relative to the $\alpha$-Fe foil at room temperature. The M\"ossbauer spectra were fitted using the commercial software package MossWinn \cite{kle16a}.

Neutron diffraction measurements were performed on single crystals of \ckfmna{} with $x$~=0.012(2), 0.017(1) and 0.036(2) with masses of 9.3(1), 7.9(1), and 5.3(1)~mg, respectively. All neutron diffraction measurements were taken on the HB-1A FIE-TAX triple-axis spectrometer at the High Flux Isotope Reactor, Oak Ridge National Laboratory. FIE-TAX operates at a fixed incident energy of 14.7 meV using two pyrolytic graphite (PG) monochromators. Two PG filters are place before and after the second monochromator to significantly reduce higher harmonics. The beam collimators placed before the monochromator, between the monochromator and sample, between the sample and analyzer, and between the analyzer and detector were $40^{\prime}-40^{\prime}-\text{S}-40^{\prime}-80^{\prime}$, respectively. Samples were sealed in an Al can containing He exchange gas which was then attached to the cold head of a closed-cycle He refrigerator. Scattering data are described using reciprocal lattice units of $h$, $k$, and $l$ for the tetragonal unit cell of $\text{CaKFe}_4\text{As}_4$. The samples were aligned with their $(H~H~L)$ planes coincident with the spectrometer's horizontal scattering plane.

\section{Results and Analyses}
\subsection{$^{75}$As NMR spectra\label{NMR}}

\begin{figure}[]
	\includegraphics[width=\columnwidth]{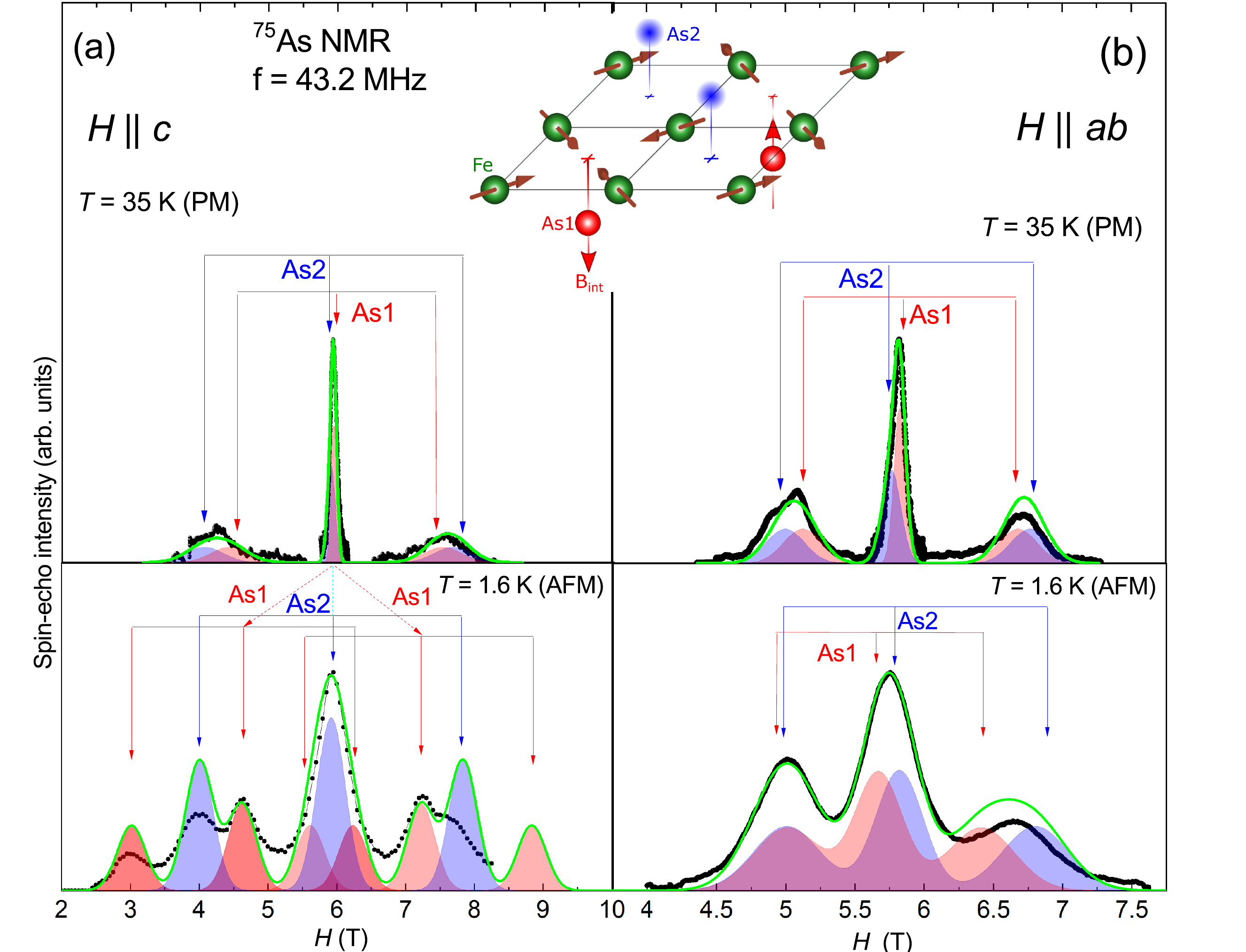} 
	\caption{{
Typical field-swept $^{75}$As-NMR spectra (black points) of CaK(Fe$_{0.964}$Mn$_{0.036}$)$_4$As$_4$ ($T_{\rm N}$ = 32 K) measured at $f$ = 43.2 MHz in the paramagnetic (top) and antiferromagnetic ordered (bottom) states, for $\bm{H}$ $\parallel$ $\bm{c}$ axis (a) and $\bm{H}$ $\parallel$ $\bm{ab}$ plane (b).  
The red and blue arrows show the calculated position of each NMR line for As1 and As2 sites, respectively, and the red and blue hatched areas show the simulated  $^{75}$As-NMR spectra for the As1 and As2 sites, respectively, with appropriate broadenings. 
  The green curves show the sum of the simulated NMR spectra for As1 and As2.    
The inset shows the sketch of the hedgehog spin-vortex crystal spin structure on an Fe-As layer. The burgundy colored arrows represent the magnetic moments at the Fe sites and the red arrows represent the internal magnetic induction $B_{\rm int}$ at the As1 site. Note  $B_{\rm int}$ is zero at the As2 site.}
}
	\label{fig:NMR}
\end{figure}

Figures \ref{fig:NMR}(a) and \ref{fig:NMR}(b) show the field-swept $^{75}$As-NMR spectra of CaK(Fe$_{0.964}$Mn$_{0.036}$)$_4$As$_4$ in both the paramagnetic state (top) and the antiferromagnetic ordered state (bottom) for two magnetic field directions, $\bm{H}\parallel\bm{c}$ axis and $\bm{H}\parallel\bm{ab}$ plane, respectively. The typical spectrum for a nucleus with spin $I=3/2$ with Zeeman and quadrupolar interactions can be described by a nuclear spin Hamiltonian ${\cal{H}}=-\gamma\hbar(1+K)HI_z+\tfrac{h\nu_Q}{6}(3I_z^2-I^2)$, where $H$ is the external field, $\hbar$ is  Planck's constant divided by $\pi$, $K$ is the Knight shift, and $\nu_{\rm Q}$ is the nuclear quadrupole frequency. The nuclear quadrupole frequency for an $I=3/2$ nuclei is given by $\nu_{\rm Q} = e^2QV_{\rm ZZ}/2h$, where $Q$ is the nuclear quadrupole moment and $V_{\rm ZZ}$ is the electric field gradient at the As site. When the Zeeman interaction is greater than the quadrupolar interaction, this Hamiltonian produces a spectrum with a sharp central transition line flanked by one satellite peak on either side. 

As observed in  $^{75}$As NMR spectra measurements in pure and Ni-doped \ckfa{} \cite{Cui2017,Ding2017,Meier2018,Ding2018,Ding2019}, the two sets of $I$ = $\frac{3}{2}$ quadrupole split lines, corresponding to the two inequivalent As sites in the paramagnetic state, are observed as shown in Fig.~\ref{fig:NMR}. The NMR lines with a larger $\nu_{\rm Q}$ of  $\sim$13 MHz (shown by blue lines in Fig.~\ref{fig:NMR}) can be assigned to the As2 site closer to the K layers, and the other lines with a smaller $\nu_{\rm Q}$  of $\sim$11 MHz (red lines) are attributed to the As1 site close to the Ca layers in CaK(Fe$_{0.964}$Mn$_{0.036}$)$_4$As$_4$, based on similar analysis used in pure \ckfa{}\cite{Cui2017}. The separation of the two As NMR lines is not as clear as those in pure and Ni-doped \ckfa{}, which may suggest that Mn substitution creates more disorder in \ckfa{}, at least as inferred by NMR measurements of the internal field at the $^{75}$As sites .
 
When $T$ is lowered below $T_{\rm N}$, for $\bm{H}$ $\parallel$ $\bm{c}$ axis, each line of the NMR spectra becomes broad and the observed spectra displays a more complicated shape as shown in the bottom panel of Fig.~\ref{fig:NMR}(a), which is similar to the NMR spectra  in the magnetically ordered state for the Ni-doped \ckfa{} \cite{Ding2017,Meier2018,Ding2019}.
The observed spectra in the AFM state can be well explained by the superposition of NMR spectra from two As sites: with the splitting of each of the NMR lines due to an internal magnetic induction $B_{\rm int}$ for the As1 site (six lines) and  no splitting of the NMR lines for the As2 site (three lines).
It is noted that the unambiguous peak assignments of the complex spectrum in the AFM state is possible by taking  the different  $\nu_{\rm Q}$ values for each As site into consideration.
When $\textbf{H}$ is applied parallel to the $\textbf{ab}$ plane, in contrast, only one set of the As1 NMR lines and one set of the As2 NMR lines have been observed as shown in the bottom panel of Fig.~\ref{fig:NMR} (b).

The difference in the NMR spectra for the different magnetic field directions can be simply explained by considering the direction of $B_{\rm int}$ for the As sites. The effective magnetic induction ${\bf B}_{\rm eff}$ is given by the vector sum of ${\bf B}_{\rm int}$ at the nucleus site and ${\bf H}$, i.e., $|$$\bf{B}_{\rm eff}$$|$ = $|$$\bf{B}_{\rm int}$ + $\bf{H}$$|$. Therefore, when $\textbf{B}_{\rm int}$ is parallel or antiparallel to $\textbf{H}$,  $B_{\rm eff}$ = $H \pm B_{\rm int}$ and a splitting of each line is expected. On the other hand, when $\textbf{H}$ $\perp$ $ \textbf{B}_{\rm int}$, no splitting of the line but a slight shift is expected since  $B_{\rm eff}$ is expressed by $B_{\rm eff}$  =  $\sqrt {H^2+B_{\rm int}^2}$. Thus, the splitting of the resonance lines at the As1 site only for $\textbf{H}$ $\parallel$ $\textbf{c}$ clearly shows that $\textbf{B}_{\rm int}$  at the As1 site is oriented along the $c$ axis. The $\textbf{B}_{\rm int}$ estimated from the splitting of the As1 central line  is around 1.30 T at 1.6 K, which is very close to the value of 1.35 T reported in CaK(Fe$_{0.951}$Ni$_{0.049}$)$_4$As$_4$ at low temperatures \cite{Ding2017,Meier2018}. The fact that neither the clear splitting nor the shift of the resonance lines has been observed for the As2 site below $T_{\rm N}$ for both $\textbf{H}$ directions indicates the net $B_{\rm int}$ at the As2 sites is zero.  

The hyperfine field pattern is only consistent with the hSVC state shown in the inset of Fig\ \ref{fig:NMR}. The line broadening in the magnetic ordered state indicates that $B_{\rm int}$ is slightly distributed probably originating from the distributions of the Fe ordered moments.

\subsection{hSVC order from neutron diffraction \label{ND}}

In order to confirm the hSVC magnetic order and study details of the magnetic structure in \ckfmna{}, we performed single-crystal neutron diffraction measurements on samples with $x$~=~0.012, 0.017 and 0.036. Features reminiscent of AFM transitions have been previously observed for $x$~=~0.017 and 0.036 in magnetization measurements. Temperature dependent magnetization, specific heat, and resistance measurements for $x$~=~0.017 and 0.036 have determined the transition temperatures of $T_\mathrm{N} = 26.2(8)$ and 32.4(9)~K, respectively\cite{Mingyu_2021}. Bulk measurements did not observe an AFM transition for $x$~=~0.012\cite{Mingyu_2021}

\begin{figure}[]
	\centering
	\includegraphics[width=\columnwidth]{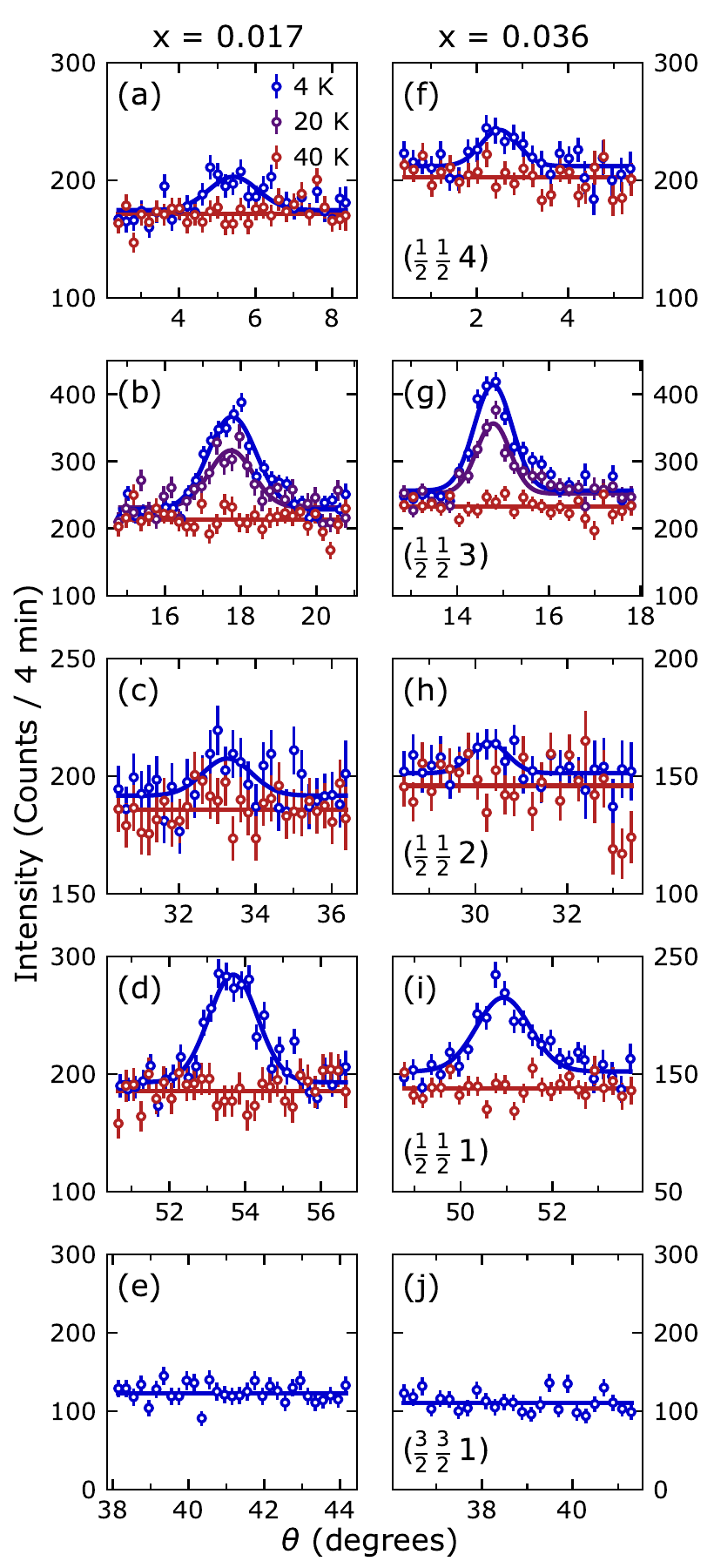}
	\caption{Rocking scans ($\theta$) of magnetic Bragg peaks of $\text{CaK(Fe}_{\text{1-}x}\text{Mn}_x\text{)}_4\text{As}_4$ measured above $T_\text{N}$ at 40~K in red, below $T_\text{N}$ at 4~K in blue and at 20~K for $(\frac{1}{2}\,\frac{1}{2}\,3)$ in purple. The data are normalized to 240 mcu which corresponds to 4 minutes of counting time. (a-e) are data for composition $x = 0.017$, and (f-j) are data for composition $x = 0.036$. Starting from the top, (a, f), magnetic Bragg peaks shown are for $(\frac{1}{2}\,\frac{1}{2}\,4), (\frac{1}{2}\,\frac{1}{2}\,3), (\frac{1}{2}\,\frac{1}{2}\,2), (\frac{1}{2}\,\frac{1}{2}\,1), (\frac{3}{2}\,\frac{3}{2}\,1)$.}
	\label{MBP}
\end{figure}

\begin{figure}[]
		\centering
		\includegraphics[width=\columnwidth]{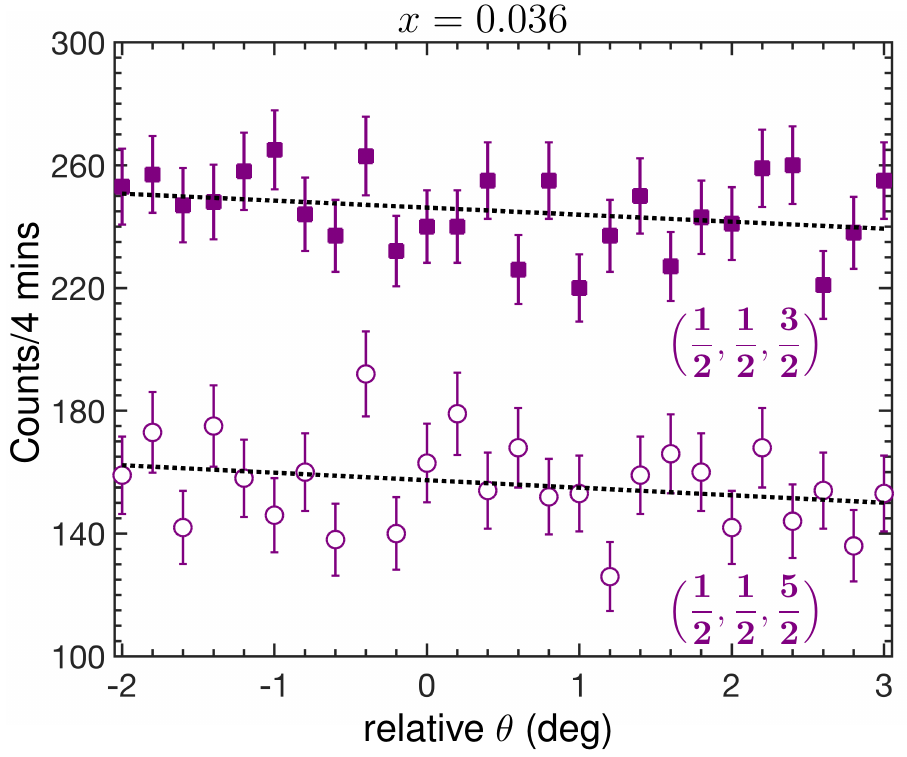}
		\caption{Rocking scans of \ckfmna{}, $x = 0.036$ at AFM Bragg peak positions with half integer L and measured at $4$~K~$< T_\text{N}$. The data are normalized to 240 mcu which corresponds to 4 minutes of counting time.}
		\label{RSHIL}
\end{figure}

Our neutron diffraction measurements (Fig.~\ref{MBP}) show magnetic Bragg peaks at the commensurate position $(\frac{1}{2}\,\frac{1}{2}\,L)$, $L=1,2,3,4$ and no magnetic Bragg peaks at $(\frac{3}{2}\,\frac{3}{2}\,1)$, which is consistent with hSVC magnetic order as seen in Ni-doped 1144. The rocking scans ($\theta$) through $(H\,H\,L)$ Bragg peaks shown in Fig.~\ref{MBP} for $x~=~0.017$ and $0.036$ single crystals at temperatures above and below the N\'eel temperature $T_\text{N}$. One can clearly see magnetic Bragg peaks appearing below $T_\mathrm{N}$. Furthermore, we do not observe peaks at half-integer $L$ as illustrated in Fig.~\ref{RSHIL}. Magnetic Bragg peaks with half-integer $H$ and integer $L$ in the $(H\,H\,L)$-plane are consistent with the doubling of the magnetic unit cell in the $\textbf{ab}$ plane but no change from the chemical unit cell along the $\textbf{c}$ direction. All these observed features are consistent with the hSVC order in \ckfna{}.

\begin{figure}[]
	\centering
	\includegraphics[width=\columnwidth]{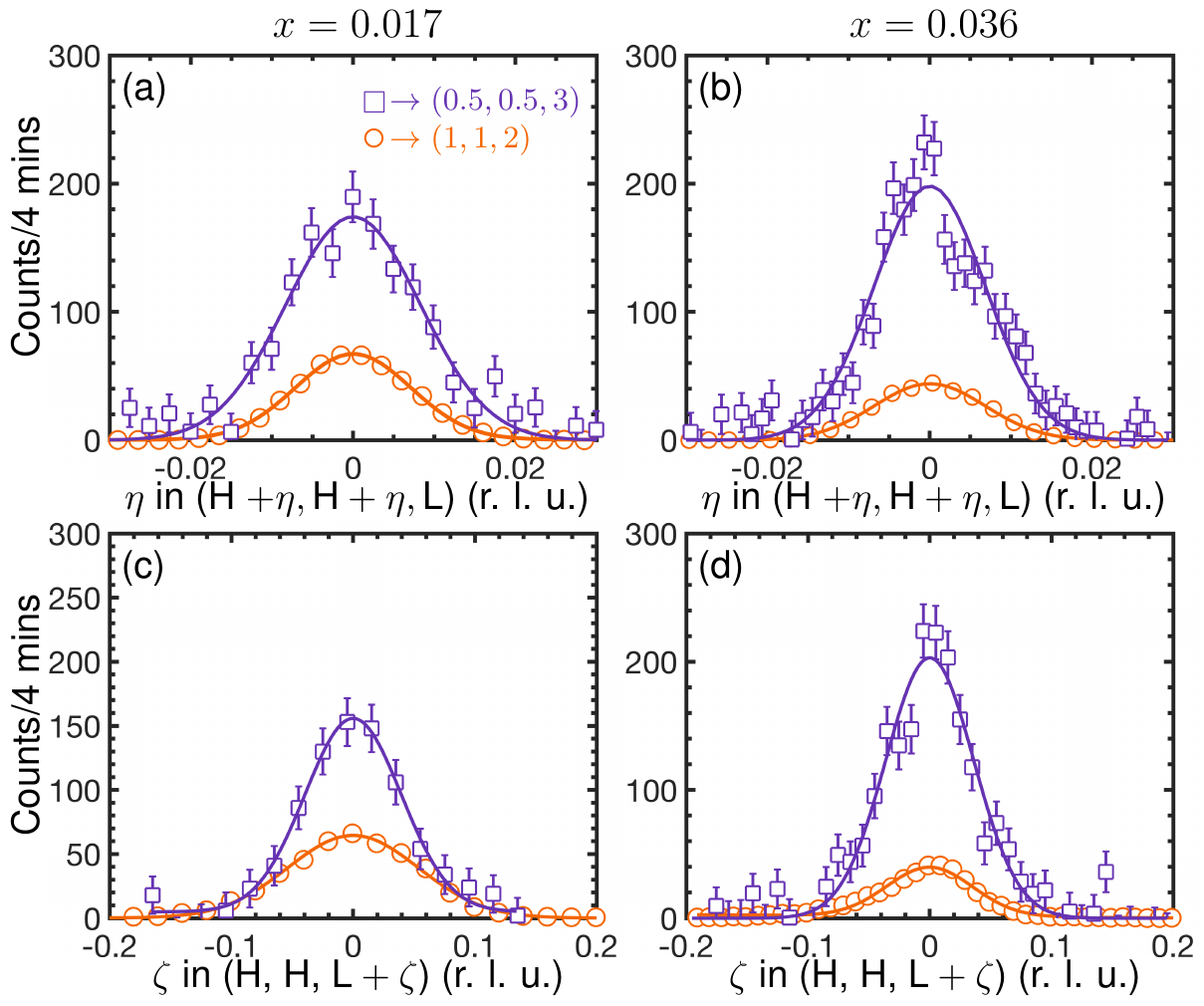}
	\caption{Neutron diffraction scans along $[H\,H\,0]$ and along the $[0\,0\,L]$ directions through the $(\frac{1}{2}\,\frac{1}{2}\,3)$ AFM Bragg peak, purple open square, for \ckfmna\ $x = 0.017$ and $0.036$ at $4$~K $< T_\mathrm{N}$. Similar scans through the $(1\,1\,2)$ Bragg peak, orange open circle, characterizing the chemical structure and the resolution conditions are shown for comparison with the intensity divided by a factor of 3000. Solid lines are Gaussian fits.}
	\label{WMN}
\end{figure}

To determine the magnetic correlation length $\xi$ in the \textbf{ab}-\text{plane} ($\xi_{ab}$) and along the $\textbf{c}$ direction ($\xi_{c}$) we fit the magnetic Bragg peak $(\frac{1}{2}\,\frac{1}{2}\,3)$ along the $[H\,H\,0]$ and $[0\,0\,L]$ direction with a Gaussian lineshapes as shown in Fig.~\ref{WMN}. For both samples $\xi_{ab}$ and $\xi_{c}$ are determined using the fullwidth, $\kappa$, of the fits and are $\approx 90$~Å and $ 60$~Å, respectively. Whereas the correlation lengths are nearly 3 times smaller than for \ckfna{}\cite{Kreyssig1144_2018}, in both cases the magnetic order is long-range since the width of both the magnetic and corresponding nuclear Bragg peaks are similar. For example, Fig.~\ref{WMN} shows the width of both the magnetic and the $(1\,1\,2)$ Bragg peaks are similar. Also, both the correlation lengths are still fairly large. The much broader nuclear and magnetic Bragg peaks in \ckfmna{} compared to \ckfna{} may be due to additional disorder present in the Mn-doped samples. Similar broadening effects were also observed and discussed in NMR measurements in Section.~\Ref{NMR}. Such disorder would be consistent with the extremely narrow range of temperatures open for growth of these crystals in comparison to the pure or Ni and Co-doped \ckfa, as described in ref.~\cite{Mingyu_2021}

 Regardless of the similarities of the magnetic Bragg peaks discussed above for \ckfmna\ and \ckfna, hSVC order is not the only possible magnetic order that can be associated with the magnetism of \ckfmna. The half integer \textit{H} and integer \textit{L} magnetic Bragg peaks are commonly associated with SSDW order in IBSC \cite{paglione2010high,lumsden2010magnetism,dai2015antiferromagnetic,goldman2008lattice}, but NMR results discussed in section.~\ref{NMR} clearly rule out SSDW order. This is not surprising since the As1 and As2 sites are inequivalent reducing the local symmetry around the Fe site from tetragonal as observed within 122-compounds to orthorhombic within the 1144 compounds. The suppression of stripe-type magnetic order is then natural when considering symmetry along with spin-orbit coupling and magneto-crystalline anisotropy, since otherwise the magnetic moment would need to lie within an arbitrary direction as can be seen in Fig.~\ref{CSMO}b. Instead, magnetic order with a spin motif which preserves tetragonal symmetry is favorable, such as the various SVC motifs \cite{o2017stabilizing,fernandes2014drives,cvetkovic2013space}.
 
 For the 1144 compounds three different possible SVC motifs within the Fe plane have been discussed and these phases can be distinguished by the relative orientation of the magnetic moment, $\bm{\mu}$, to the propagation vector: (i) $\bm{\mu}_{\textbf{ab}} \parallel \bm{\tau}_{\textbf{ab}}$, (ii) $\bm{\mu}_{\textbf{ab}} \perp \bm{\tau}_{\textbf{ab}}$, and (iii) $\bm{\mu} \perp \textbf{ab}$ \cite{Meier2018, cvetkovic2013space,o2017stabilizing, christensen2015spin,fernandes2016vestigial}. Within the literature these SVC orders have been colloquially described as (i) hSVC order, (ii) loops SVC order, and (iii) spin charge-density wave (SCDW) order as illustrated in Fig.~1 of Ref.~\cite{Meier2018}. These three SVC motifs become six AFM structures since each can be stacked FM or AFM with their nearest-neighbor along \textbf{c}. 

\begin{table*}[]
	\caption{\label{Tab:diff_representations}  AFM Bragg peak integrated intensity measured on $\text{CaK(Fe}_{\text{1-}x}\text{Mn}_x\text{)}_4\text{As}_4$ single crystals, and corresponding calculated intensity for hSVC, Loops SVC, and SCDW magnetic order. Intensities are given in arbitrary units and normalized to the intensities of several chemical Bragg peaks. Calculated values are for 0.38~$\mu_\text{B}$ at each Fe/Mn site as described in Ref.~\cite{Kreyssig1144_2018}} 
	\begin{ruledtabular}
		\begin{tabular}{P{0.7in}P{0.7in}P{0.7in}P{0.7in}P{0.7in}P{0.7in}P{0.7in}P{0.7in}P{0.7in}}
			\centering
			%		\multicolumn{1}{c}{ } & \multicolumn{2}{c}{Measurements}  & \multicolumn{6}{c}{Calculation}\\
			%		\cmidrule(lr){2-3}\cmidrule(lr){4-9}
			AFM \newline Bragg peak & $x$~= 0.017 \newline $T$~=~4~K & $x$~= 0.036 \newline $T$~=~4~K & \multicolumn{2}{P{1.5in}}{hSVC in plane: \newline $\mu_i~\parallel~\tau_i$} & \multicolumn{2}{P{1.5in}}{ Loops SVC in plane: \newline $\mu_i~\perp~\tau_i$} & \multicolumn{2}{P{1.5in}}{SCDW in plane: \newline $\mu_i~\parallel~\tau_i$} \\
			%	\cmidrule(lr){4-5}\cmidrule(lr){6-7}\cmidrule(lr){8-9}
			&  &  & AFM along c & FM along c & AFM along c & FM along c & AFM along c & FM along c\\
			\hline \\[-5pt]
			$(\frac{1}{2}\,\frac{1}{2}\,4)$ & 21 & 32 & 19 & 80 & 26 & 107 & 13 & 54\\[5pt]
			$(\frac{1}{2}\,\frac{1}{2}\,3)$ & 104 & 167 & 130 & 17 & 207 & 26 & 155 & 20\\[5pt]
			$(\frac{1}{2}\,\frac{1}{2}\,2)$ & 9 & 5 & 9 & 164 & 21 & 384 & 24 & 441\\[5pt]
			$(\frac{1}{2}\,\frac{1}{2}\,1)$ & 69 & 94 & 99 & 1 & 634 & 8 & 1071 & 14\\[5pt]
			$(\frac{3}{2}\,\frac{3}{2}\,1)$ & $<2$ & $<2$ & 1 & 0.1 & 32 & 0.4 & 63 & 1
		\end{tabular}
	\end{ruledtabular}
\end{table*}

To firmly establish the correct SVC motif for \ckfmna{}, we compared the measured integrated intensities of several $(\frac{1}{2}\,\frac{1}{2}\,L)$ peaks and the $(\frac{3}{2}\,\frac{3}{2}\,1)$ peak to their calculated values using FULLPROF~\cite{rodriguez1993fullprof}. The results for all six possible SVC motifs are listed in Table \ref{Tab:diff_representations}. Clearly, for compositions $x$~=~0.017 and 0.036, only the hSVC order with AFM stacking along the $\bm{c}$ direction is consistent with the available data, i.e.~the maximum intensity at $(\frac{1}{2}\,\frac{1}{2}\, 3)$ and minimum at $(\frac{3}{2}\,\frac{3}{2}\, 1)$. By fitting the measured intensities of each Bragg peak to the corresponding calculated values, we determined the magnetic moment at 4~K on each transition metal site to be $\mu$ = 0.33(5)~$\mu_{\text{B}}$ and 0.40(5)~$\mu_{\text{B}}$ for $x$ = 0.017 and 0.036, respectively. The large Mn moment of $\approx5$~$\mu_{\text{B}}$ can account for approximately up to 0.085~$\mu_{\text{B}}$ and 0.18~$\mu_{\text{B}}$ of the ordered magnetic moment for $x$ = 0.017 and 0.036, respectively. We cannot identify the origin of the ordered magnetic moment with the present data, but the ordered moment is similar to the approximately 0.4~$\mu_{\text{B}}$ found for \ckfna{} \cite{Kreyssig1144_2018}. 

These results show that the magnetic structure of \ckfmna{} is the same as that of \ckfna{}. This means that the magnetic space group of each is $P_{C}4/mbm$ (BNS) for the AFM unit cell or $P_{P}4^{\prime}/mmm^{\prime}$ (OG) for the chemical unit cell, as discussed in Ref.~\cite{Kreyssig1144_2018}. This is a $\text{two-}\tau$ AFM structure with propagation vectors $\bm{\tau}_\mathrm{1} = (\frac{1}{2}\,\frac{1}{2}\,1)$ and $\bm{\tau}_\mathrm{2} = (-\frac{1}{2}\,\frac{1}{2}\,1)$ in reciprocal lattice units, and $(\pi\,0)$ and $(0\, \pi)$ in the one-Fe Brillouin zone notation \cite{paglione2010high,Meier2018,Kreyssig1144_2018}.

\begin{figure}[]
	\centering
	\includegraphics[width=\columnwidth]{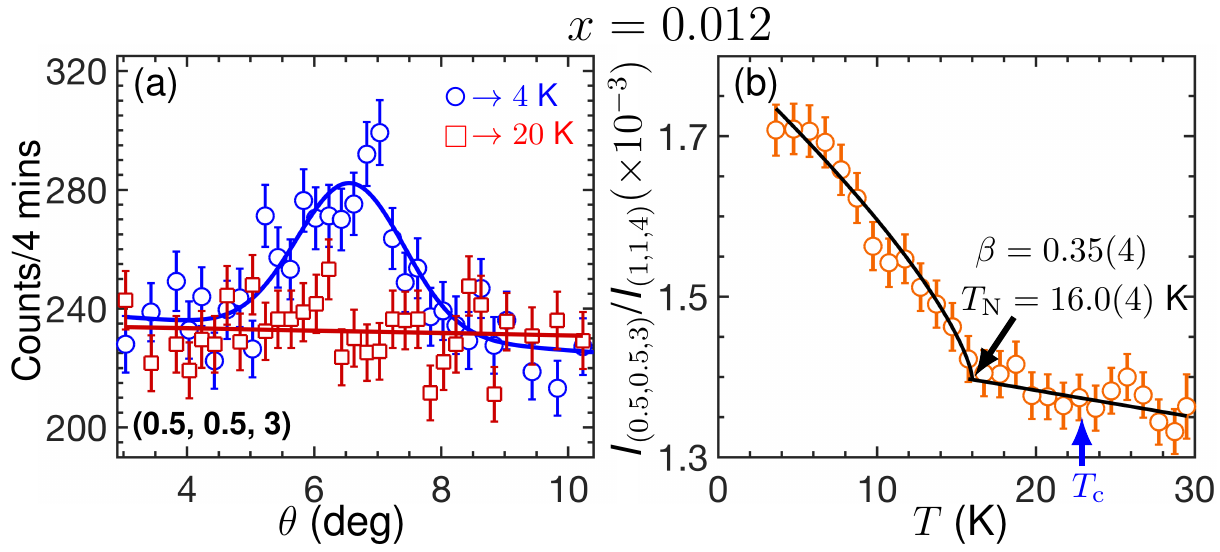}
	\caption{(a) Rocking scans of \ckfmna, $x = 0.012$ at (0.5, 0.5, 3) AFM Bragg peak and at temperatures above and below $T_\mathrm{N}$. (b) Temperature dependence of $(0.5, 0.5,3)$ AFM Bragg peak normalized to the $(1\,1\,4)$ peak intensity illustrating the magnetic order below $T_\mathrm{c}$. $T_\mathrm{N}$ and critical exponent $\beta$ obtained from a critical exponent fit is given in the figure.}
	\label{x_0p012}
\end{figure}

Whereas the $x_\text{Mn}=0.017$ and 0.036 samples had $T_\mathrm{N}>T_\mathrm{c}$ values that can be inferred from bulk measurements (see Fig.~\ref{PD} above) we  were able to measure $T_\mathrm{N}<T_\mathrm{c}$ for a $x_\text{Mn}=0.012$ sample. This is a very important data point, since together with the M\"ossbauer spectroscopy data discussed in the next section it allows us to determine the slope of the $T_\mathrm{N}$ line under the SC dome. Figure~\ref{x_0p012}(a) shows rocking scans through the AFM $(\frac{1}{2}\,\frac{1}{2}\,3)$ Bragg peak for the $x = 0.012$ sample and its absence by 20~K. The temperature dependence of the normalized intensity, in Fig.~\ref{x_0p012}(b) clearly shows that $T_\mathrm{N}=16.0(4)~\text{K} < T_\mathrm{c}$. It is also apparent that the magnetic phase transition is still second-order. The $T_\mathrm{N}$ determined for the $x=0.012$ sample combined with a comparable data point from M\"ossbauer spectroscopy (below), clearly indicate that the slope of the $T_\mathrm{N}(x)$ line changes dramatically below $T_\mathrm{c}$, suggesting the presence of strong competition between SC and AFM states as outlined in Refs.~\cite{maple1976superconductivity, machida1981spin, zhang2002competing, Fernandes2010unconventional,wolowiec2015conventional} \textcolor{red}.

\begin{figure}[]
	\centering
	\includegraphics[width=\columnwidth]{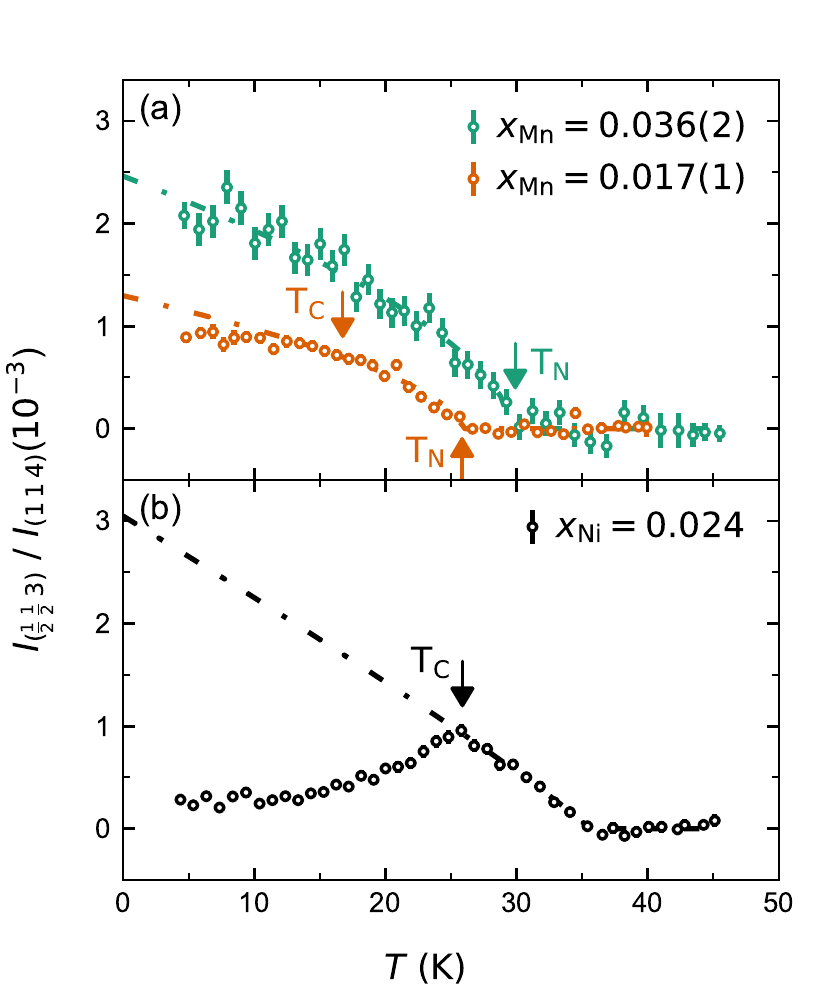}
	\caption{Temperature dependence of $(\frac{1}{2}\,\frac{1}{2}\,3)$ AFM Bragg peak normalized to the $(1\,1\,4)$ peak intensity for several compositions of $\text{CaK(Fe}_{\text{1-}x}T_x\text{)}_4\text{As}_4$ $T$~=~Mn, Ni. Dotted lines are a guide to the eye generated by a fit of the order parameter below $T_\text{N}$ for each composition. Arrows are used to indicate $T_\text{c}$ and placed based on magnetization data for $T$~=~Mn\cite{Mingyu_2021}, and by the kink in the intensity for $T$~=~Ni.}
	\label{OrderParam}
\end{figure}

Figure~\ref{OrderParam} shows a detailed temperature dependence of the intensity of the magnetic Bragg peak $(\frac{1}{2}\,\frac{1}{2}\,3)$ for \ckfmna\ $x = 0.017$ and $0.036$ in (a) compared to \ckfna\ $x_\mathrm{Ni} = 0.024$ shown in (b). The order parameter is well described by a critical exponent fit for a second-order transition with a sharp onset. The critical exponent fit gives a N\'eel temperature $T_\text{N}$~=~26.1(4) and 30.2(2)~K for $x_\text{Mn}$~=~0.017 and 0.036, respectively. The $T_\mathrm{N}$ values for each composition are in agreement with bulk measurements as shown in Fig.~\ref{PD}. Each power law fit resulted in critical exponents $\beta \approx 0.34$ for both of the samples.

The suppression of the ordered magnetic moment below $T_\mathrm{c}$ has been considered a straightforward signature of competition between coexisting magnetism and SC for the same electrons in IBSC \cite{Kreyssig1144_2018}. On comparing the data in Fig.~\ref{OrderParam}(a) to those shown in Fig.~\ref{OrderParam}(b), we see that the decrease in magnetic intensity below $T_\mathrm{c}$ observed in the Ni-doped sample is not resolvable in the superconducting $x = 0.017$ sample. This feature often fades out when $T_\mathrm{c}$ is far below $T_\mathrm{N}$, as can be seen in Ref.~\cite{Kreyssig1144_2018} for $x_\mathrm{Ni} = 0.051$ where $T_\mathrm{c}/T_\mathrm{N}~\tildeapprox{}~0.18$. For the Ni-doped case this feature is robust for $x_\mathrm{Ni} = 0.024$ as shown in Fig.~\ref{OrderParam}(b), and $x_\mathrm{Ni} = 0.033$ as shown in Ref.~\cite{Kreyssig1144_2018} , where the ratio $T_\mathrm{c}/T_\mathrm{N}$ is approximately 0.72 and 0.50, respectively. The ratio of $T_\mathrm{c}/T_\mathrm{N}$ for $x = 0.017$ is approximately 0.65, which is comparable to the values of $T_\mathrm{c}/T_\mathrm{N}$ where suppression of the magnetic moment is observed below $T_\mathrm{c}$ for Ni-doped \ckfa{}.

\subsection{M\"ossbauer spectroscopy}

\begin{figure}[]
	\centering
	\includegraphics[width=\columnwidth]{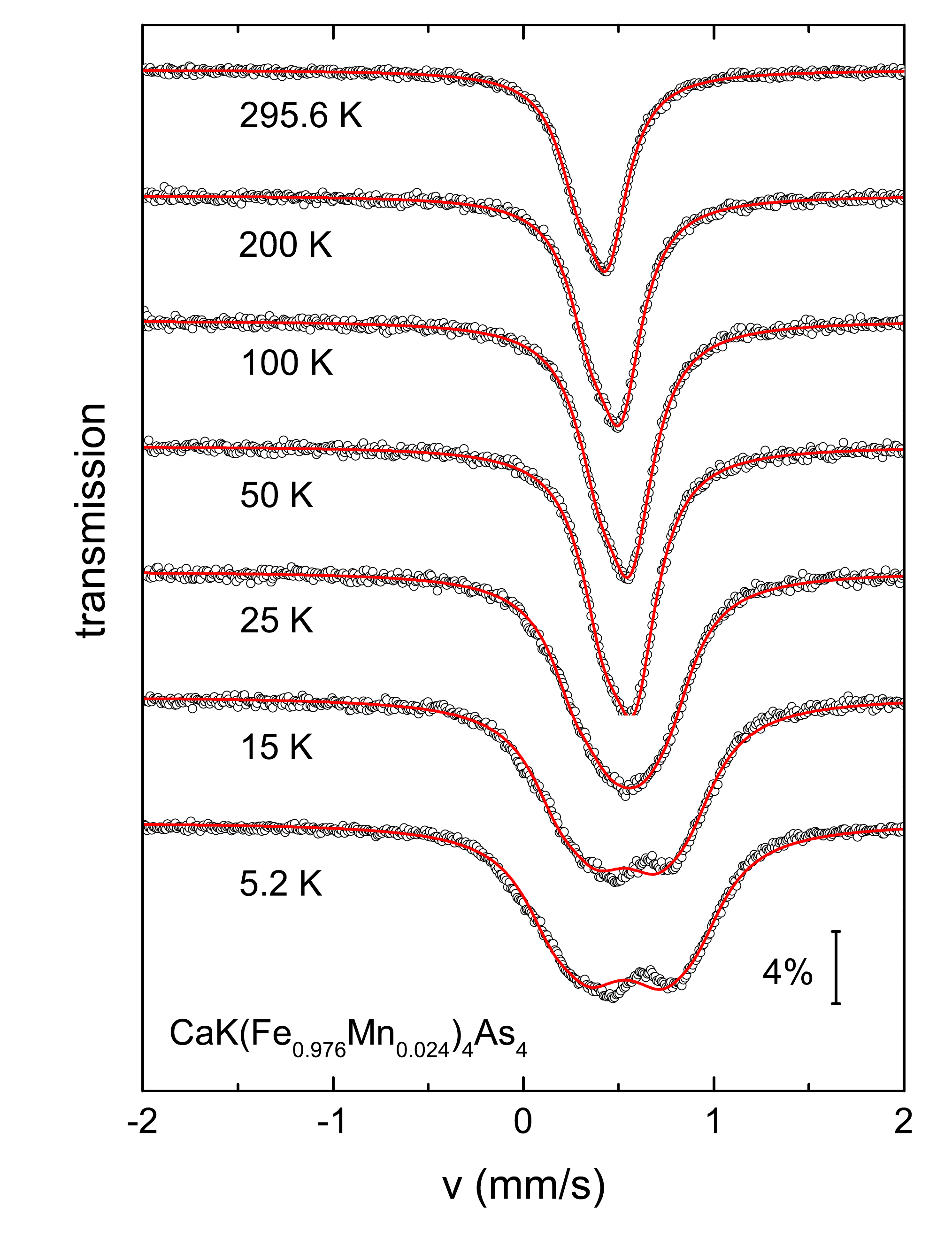}
	\caption{$^{57}$Fe M\"ossbauer spectra of CaK(Fe$_{0.976}$Mn$_{0.024}$ )$_4$As$_4$ sample at selected temperatures. Symbols: data, lines: fits; see text for details.}
	\label{spectra}
\end{figure}

The $^{57}$Fe M\"ossbauer spectra at different temperatures from room temperature down to $\sim 5$ K were collected for five Mn concentrations, $x = 0.010, 0.012, 0.017, 0.020$, and $0.024$, of CaK(Fe$_{1-x}$Mn$_{x}$)$_4$As$_4$.  A subset of M\"ossbauer spectra for CaK(Fe$_{0.976}$Mn$_{0.024}$)$_4$As$_4$ sample is shown in Fig. \ref{spectra}.  Similar behavior (with somewhat different characteristic temperatures and hyperfine parameters) was observed for four other samples. The absorption lines for temperatures between 50 and 295.6~K, in the paramagnetic state,  are asymmetric, suggesting that each spectrum is a quadrupole split doublet with a rather small value of the quadrupole splitting (QS). At low temperatures the spectra broaden and change their shape. These low temperature data can be fit with a magnetic sextet. The full Hamiltonian approach [“Mixed M + Q Static Hamiltonian (Mosaic)”] model in the MossWinn \cite{kle16a} software package was used to analyze these spectra. To limit the number of fitting parameters, the moments were fixed to be in the $\bm{ab}$ plane (as follows from the NMR and neutron results discussed above) and the QS values in ordered state were fixed to the value observed in the paramagnetic state just above the magnetic ordering transition.

\begin{figure}[]
	\centering
	\includegraphics[width=\columnwidth]{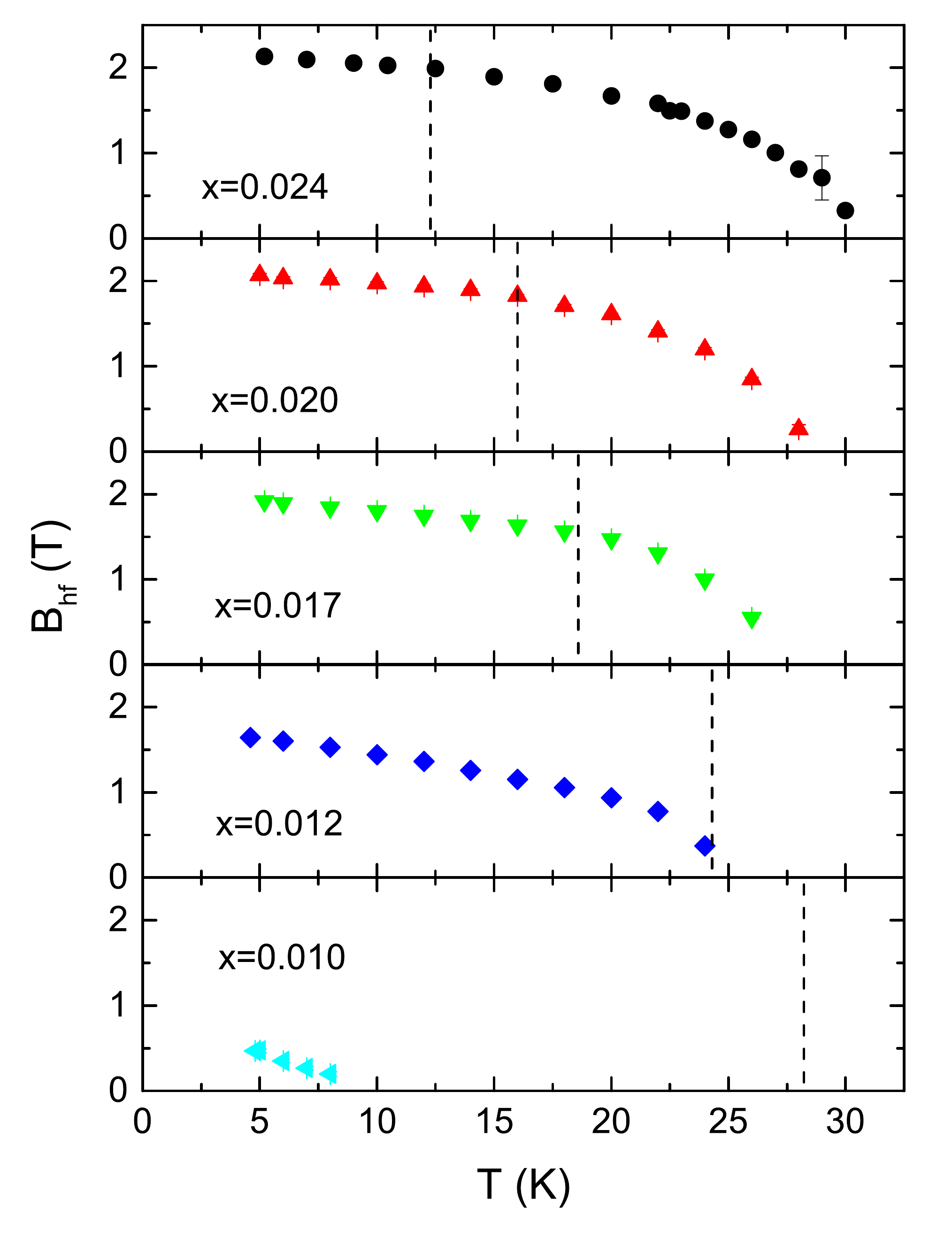}
	\caption{Temperature dependence of the hyperfine field for five CaK(Fe$_{1-x}$Mn$_{x}$ )$_4$As$_4$ sample. Vertical dashed lines mark superconducting transition temperatures.}
	\label{Bhf}
\end{figure}

The temperature dependence of the hyperfine field, $B_{\text{hf}}$, on the $^{57}$Fe sites for all five samples is summarized in Fig. \ref{Bhf}. Similar to the neutron data, little to no features corresponding to the suppression of the magnetic order below $T_\mathrm{c}$ are observed in $x\geq0.017$ samples, where $T_\mathrm{c} < T_\mathrm{N}$. In addition, the magnetic phase transition for all five concentrations are second-order. Also, $T_\mathrm{N} < T_\mathrm{c}$ is observed for lower concentration samples $x = 0.012$. The $x=0.012$ M\"ossbauer sample has a clear $T_\text{N}$ value of 24.3~K as can be seen from the $B_{hf}$ data shown in Fig.~\ref{Bhf}. This value of $T_\text{N}$ is different from the value found for a sample with the same nominal $x$ value by neutron diffraction [17.1~K from Fig.~\ref{x_0p012}(b)], but this difference is consistent with the very steep slope of $T_\text{N}$ shown in Fig.~\ref{PD}. Small differences in concentration will lead to large differences in the $T_\text{N}$. Whereas the M\"ossbauer data was collected on an ensemble of samples, the neutron diffraction data was collected on a single crystal.

\section{Discussion and Summary}

In Fig.~\ref{PD} we combine our results of magnetic transition temperatures with the $T-x$ phase diagram presented in Ref.~\cite{Mingyu_2021}. Three different low-temperature regions are found in the phase diagram labeled as SC, SC + AFM (coexistence region) and AFM. The sudden deviation of the magnetic phase line when $T_\mathrm{N}< T_\mathrm{c}$ compared to the higher temperature, $T_\mathrm{N}< T_\mathrm{c}$, magnetic phase line indicates that the AFM phase is clearly suppressed by SC. In addition, SC is clearly suppressed with increasing Mn concentration. Perhaps the most striking part of the $T-x$ phase diagram is the clear delineation of $T_\mathrm{N}(x)$ for $T_\mathrm{N}< T_\mathrm{c}$. We can clearly identify a critical doping value, $x_\text{c}\tildeapprox{}0.01$. This concentration corresponds to a quantum phase transition, which is likely a quantum critical point associated with the onset of AFM order under the SC dome.

Suppression of SC and the appearance of AFM order with substitution of Fe by Mn in optimally-doped superconducting IBSCs has been observed in compounds such as \bkfmna{} \cite{Cheng_2010} and \lfmaf{} \cite{Hammerath_2014}. Compared to Ni or Co doping the Mn acts as a much larger, moment-bearing impurity. These local magnetic impurity scattering centers act as pair breakers thereby suppressing the SC \cite{abrikosov1960contribution}. Also, local magnetic Mn impurities have been observed to promote tetragonal AFM order via cooperative coupling with the conduction electrons in Mn-doped \bfa{} \cite{Inosov_2013,Gastiasoro_2014}.

The above mechanism of magnetic pair-breaking can explain several features in the \ckfmna{} series as well. Where \ckfa{} is an optimally hole-doped member, \ckfna{} with electron doping moves the system towards the AFM region of the phase diagram, and \ckfmna{} is the Mn-doped counterpart. Therefore, local magnetic scattering centers originating from Mn substitution can break superconducting pairs and suppress SC in \ckfmna{}. SC being suppressed by local magnetic scattering centers is supported by the following observations. There is a sharp decrease of $T_{\text{c}}$ with Mn substitution compared to Ni and Co substitution. The localization of the charge carriers is consistent with the large resistivity observed in Ref.~\cite{Mingyu_2021}. In addition, the jump in $C_p$ at a given $T_{\text{c}}$ is significantly suppressed for Mn substitution compared to Ni substitution in \ckfa{} \cite{Mingyu_2021}.

Hence, except for the coexisting magnetic and superconducting region, the AFM order with Mn doping in 1144 is similar to other Mn-doped IBSCs. \lfmaf\ (1111) is another member where such coexistence has been observed, but only short-range AFM order coexists with SC rather than the long-range order found in our $x = 0.012$ sample\cite{Hammerath_2014}. Attempts to fit the low-temperature M\"ossbauer spectra as a sum of paramagnetic (doublet) and magnetic (sextet) components do not improve the quality of the fit, suggesting that there is no phase separation in the samples.

Another peculiar difference in \ckfmna{} is shown in Figs.~\ref{OrderParam} and \ref{Bhf} where the suppression of magnetic order below $T_\mathrm{c}$ apparent for \ckfna{} is conspicuously absent. The suppression below $T_\mathrm{c}$ is due to a decrease in the magnetic moment at the Fe site, which can be triggered by an opening of the superconducting gaps at the Fermi surface \cite{Pratt_2009, Munevar_2013}. It has previously been discussed that the suppression is consistent with the microscopic coexistence, or competition for the same electrons \cite{machida1981spin, Fernandes2010unconventional, bud2018coexistence, Kreyssig1144_2018}. We do not resolve such a change of the ordered magnetic moment in \ckfmna{}. This may be because the suppression of $T_\mathrm{c}$ by more local-moment-like Mn alters the simple picture of SC and AFM competition discussed in Ref.~\cite{bud2018coexistence} for \ckfna{}.

Using NMR, M\"ossbauer spectroscopy, and single-crystal neutron diffraction measurements, we found details of the magnetic order in \ckfmna{} series and established three low-temperature regions in the phase diagram- SC, SC + AFM (coexistence region) and AFM. We identified a critical doping value of $x_\text{c}\tildeapprox{}0.10$ which may be a quantum critical point associated with the onset of AFM order under the SC dome. We discussed the features which are similar and starkly different from \ckfna{}. Both series show suppression of SC with increasing Mn/Ni substitution and appearance of long-range commensurate hSVC order that is described by symmetry equivalent propagation vectors $(\pi\,0)$ and $(0\,\pi)$ and have AFM correlations along the $\textbf{c}$ direction. Suppression of $T_\mathrm{N}$ in the coexistence region indicates a strong competition between magnetism and SC. On the other hand, no resolvable suppression of the ordered magnetic moment below $T_\mathrm{c}$ for \ckfmna{} $x = 0.017$ suggests a change in the details of SC and AFM competition due to the introduction of magnetic impurities from the Mn substitution.
\\
\begin{acknowledgments}
	We are grateful for Dominic H. Ryan for useful discussions. Work at Ames National Laboratory was supported by the U.\,S.\ Department of Energy (DOE), Basic Energy Sciences, Division of Materials Sciences \& Engineering, under Contract No.\ DE-AC$02$-$07$CH$11358$. A portion of this research used resources at the High Flux Isotope Reactor, a U.\,S.\ DOE Office of Science User Facility operated by Oak Ridge National Laboratory. SLB is indebted to Dr. Zolt\'an Klencs\'ar for his help in modification of MossWinn code.
\end{acknowledgments}

\bibliographystyle{apsrev4-2.bst}
\bibliography{CaKFe-Mn4As4.bib}

\end{document}